\documentclass[twocolumn,english]{revtex4-2}
\usepackage[T1]{fontenc}
\usepackage[latin9]{luainputenc}
\usepackage{geometry}
\usepackage{amsmath}
\usepackage{amssymb}
\usepackage{babel}
\begin{document}
\title{Higher Derivative Muffin Tin Orbitals (HDMTO) and Higher Derivative
Koringa Khon and Rostoker (HDKKR) methods}
\author{Garry Goldstein$^{1}$}
\address{$^{1}$garrygoldsteinwinnipeg@gmail.com}
\begin{abstract}
In this work we introduce a Linearized version of the Koringa Khon
and Rostoker method (LKKR) and show it to be equivalent to the Linearized
Muffin Tin Orbitals method (LMTO). We then present higher derivative
versions of both methods, e.g. HDKKR and HDMTO and show them to be
partially distinct (not equivalent). In particular HDKKR basis set
does not have an equivalent ground state for the Khon Sham (KS) Hamiltonian
as the HDKKR basis set and has greater variational power then the
HDMTO one. Because the KS method, for Density Functional Theory (DFT),
is variational HDKKR will give better ground state energies then HDMTO.
However HDKKR is much harder to work with then HDMTO requiring much
greater computer resources so HDMTO can often be preferred.
\end{abstract}
\maketitle

\section{\protect\label{sec:Introduction}Introduction}

The choice of basis set is fundamental for an efficient solution of
a Density Functional Theory problem (DFT) - the diagonalization of
the Khon Sham (KS) Hamiltonian. The basis set must be efficient, in
that a small number of basis elements well approximate the exact Khon
Sham (KS) wavefunctions at least in the middle of the valence band.
The basis must also be simple to manipulate - with various practical
calculations associated with finding matrix element of the KS Hamiltonian
with respect to the basis easily implementable on computer. The basis
set must be transferable, that is one does not want to use a significantly
different basis for every single many body problem or equivalently
for diagonalizing every KS Hamiltonian. The Koringa Khon Rostoker
(KKR) and the Augmented plane Wave (APW) basis sets are such basis
sets, however they are deficient in that the basis elements, of both
basis sets, depend explicitly on the energy of the eigenstate, to
be computed, which leads to a self consistent calculation of the basis
set thereby increasing computational costs by easily an order of magnitude
\citep{Singh_2005,Martin_2020}. The key idea to overcome this difficulty
is due to O. K. Andersen who linearized the basis set (with the linearization
energy being chosen in the middle of the valance band) thereby obtaining
a fixed basis for each iteration of the self consistency loop for
the solution of the self consistent KS problem \citep{Andersen_1975,Singh_2005,Martin_2020,Marx_2009},
each step requires simply the diagonalization of a matrix. This lead
to the introduction of the Linearized Augmented Plane Wave (LAPW)
\citep{Andersen_1975,Singh_2005,Martin_2020} and Linearized Muffin
Tin Orbitals (LMTO) methods \citep{Skriver_1984,Wills_2010}. In both
these methods the solid is divided into a Muffin Tin (MT) sphere part
(with spheres centered around atomic nuclei) and an interstitial part
\citep{Andersen_1975,Singh_2005,Martin_2020}). In the interstitial
part the KS potential is assumed to be sufficiently smooth that plane
waves or spherical Bessel functions form a good basis for the region
while the basis set is adapted to the MT part to be the solution of
the spherically averaged KS Hamiltonian. The solution to the spherically
averaged KS Hamiltonian is chosen at some linearization energy (typically
in the middle of the valence band) and the solution $\psi_{l\mu}^{E}\left(r\right)$
as well as $\dot{\psi}_{l\mu}^{E}\left(r\right)=\frac{\partial}{\partial E}\psi_{l\mu}^{E}\left(r\right)$
- its derivative with respect to energy - are used as a basis set
inside the MT spheres. Further accuracy may be obtained for the HDLO
method where $\ddot{\psi}_{l\mu}^{E}\left(r\right)=\frac{\partial^{2}}{\partial E^{2}}\psi_{l\mu}^{E}\left(r\right)$
is used to augment the basis set \citep{Michalicek_2013,Martin_2020}.
In this work after a brief review of the usual KKR method we linearize
the KKR method to LKKR and show it to be equivalent to LMTO. We then
extend these ideas to higher derivative versions HDKKR and HDMTO and
show that HDKKR basis set has greater variational power then the HDMTO
basis set. However HDKKR is much harder to implement then HDLMTO and
requires much greater computer resources so often HDMTO is preferred.

\section{\protect\label{sec:Regular-KKR-method-1}Regular KKR method (review)}

We now write the KKR basis set wavefunctions: 
\begin{align}
 & a_{l}^{\mu}\Phi_{lm}^{1\mu}\left(E,\mathbf{r}\right)=Y_{lm}\left(\widehat{\mathbf{r-\mathbf{r}}_{\mu}}\right)\times\nonumber \\
 & \times\left\{ \begin{array}{cc}
\psi_{l\mu}^{E}\left(\left|\mathbf{r}-\mathbf{r}_{\mu}\right|\right)-b_{l}^{\mu}J_{l}^{\kappa}\left(\left|\mathbf{r}-\mathbf{r}_{\mu}\right|\right) & \left|\mathbf{r}-\mathbf{r}_{\mu}\right|\leq S^{\mu}\\
a_{l}^{\mu}K_{l}^{\kappa}\left(\left|\mathbf{r}-\mathbf{r}_{\mu}\right|\right) & \left|\mathbf{r}-\mathbf{r}_{\mu}\right|>S^{\mu}
\end{array}\right.\label{eq:KKR_basis}
\end{align}
Here then $1$ in $\Phi_{lm}^{1\mu}\left(E,\mathbf{r}\right)$ is
for later notational use. Here: 
\begin{equation}
\left[-\frac{d^{2}}{dr^{2}}+\frac{l\left(l+1\right)}{r^{2}}+\bar{V}_{KS}\left(r\right)\right]r\psi_{l\mu}^{E}\left(r\right)=Er\psi_{l\mu}^{E}\left(r\right)\label{eq:Schrodinger_equation}
\end{equation}
and $\bar{V}_{KS}\left(r\right)$ is the spherically average Khon
Sham (KS) potential. Where furthermore: 
\begin{align}
\kappa^{2} & =E\nonumber \\
K_{l}^{\kappa}\left(r\right) & =-\kappa^{l+1}\left\{ \begin{array}{cc}
n_{l}^{\kappa}\left(r\right) & \kappa^{2}>0\\
n_{l}^{\kappa}\left(r\right)-ij_{l}^{\kappa}\left(r\right) & \kappa^{2}<0
\end{array}\right.\nonumber \\
J_{l}^{\kappa}\left(r\right) & =\kappa^{-l}j_{l}^{\kappa}\left(r\right)\label{eq:Normalization_conventions}
\end{align}
Here $n_{l}$ and $j_{l}$ are the spherical Neumann and Bessel functions
respectively. Now we look for eigenfunctions of the form: 
\begin{equation}
\chi\left(E,\mathbf{k},\mathbf{r}\right)=\sum_{\mathbf{R}}\exp\left(i\mathbf{k}\cdot\mathbf{R}\right)\sum_{lm}A_{lm}^{\mu}\left(\mathbf{k},E\right)\Phi_{lm}^{1\mu}\left(E,\mathbf{r}-\mathbf{R}\right)\label{eq:Trial_wavefunction-1}
\end{equation}
Now we write:
\begin{align}
 & \sum_{\mathbf{R}+\mathbf{r}_{\nu}\neq\mathbf{r}_{\mu}}\exp\left(i\mathbf{k}\cdot\mathbf{R}\right)K_{l}^{\kappa}\left(\left|\mathbf{r}-\mathbf{R}-\mathbf{r}_{\nu}\right|\right)Y_{lm}\left(\widehat{\mathbf{r}-\mathbf{R}-\mathbf{r}_{\nu}}\right)\nonumber \\
 & =\sum_{l',m'}B_{\nu\mu l'm';lm}\left(\mathbf{k}\right)J_{l'}^{\kappa}\left(\left|\mathbf{r}-\mathbf{r}_{\mu}\right|\right)Y_{l'm'}\left(\widehat{\mathbf{r}-\mathbf{r}_{\mu}}\right)\label{eq:Structure_factors}
\end{align}
For the some structure constants $B_{l'm';lm}^{\left\{ \mathbf{r}_{\mu}\right\} }\left(\mathbf{k}\right)$
\citet{Wills_2010}. Now we wish to eliminate all tails in all basis
functions that is Bessel like wavefunction components inside all MT
spheres. As such we want for $\left|\mathbf{r}-\mathbf{r}_{\mu}\right|\leq S^{\mu}$:
\begin{align}
 & \chi\left(E,\mathbf{k},\mathbf{r}\right)\nonumber \\
 & =\sum_{lm}A_{lm}^{\mu}Y_{lm}\left(\widehat{\mathbf{r}-\mathbf{r}_{\mu}}\right)\psi_{l}^{E}\left(\left|\mathbf{r}-\mathbf{r}_{\mu}\right|\right)\label{eq:Good_terms}
\end{align}
However we have that: 
\begin{align}
 & \chi\left(E,\mathbf{k},\mathbf{r}\right)\nonumber \\
 & =\sum_{l'm'}A_{l'm'}^{\mu}\left(\mathbf{k},E\right)Y_{l'm'}\left(\widehat{\mathbf{r}-\mathbf{r}_{\mu}}\right)a_{l'}^{\mu-1}\times\nonumber \\
 & \times\left[\psi_{l'}^{E}\left(\left|\mathbf{r}-\mathbf{r}_{\mu}\right|\right)-b_{l'}^{\mu}J_{l'}^{\kappa}\left(\left|\mathbf{r}-\mathbf{r}_{\mu}\right|\right)\right]+\nonumber \\
 & +\sum_{\nu lm}A_{lm}^{\nu}\left(\mathbf{k},E\right)\sum_{l',m'}B_{\nu\mu l'm';lm}\left(\mathbf{k}\right)\times\nonumber \\
 & \times J_{l}^{\kappa}\left(\mathbf{r}-\mathbf{r}_{\mu}\right)Y_{l'm'}\left(\widehat{\mathbf{r}-\mathbf{r}_{\mu}}\right)\label{eq:All_terms}
\end{align}
Equating the Right Hand Sides of Eqs. (\ref{eq:Good_terms}) and (\ref{eq:All_terms})
we obtain the relationship 
\begin{equation}
\sum_{\mu lm}\left[B_{\nu\mu l'm';lm}\left(\mathbf{k}\right)-\frac{b_{l}^{\mu}}{a_{l}^{\mu}}\delta_{\mu\nu;ll';mm'}\right]A_{lm}^{\mu}\left(\mathbf{k},E\right)=0\label{eq:KKR_relation}
\end{equation}
Furthermore we wish for the wavefunction to be continuous and continuously
differentiable MT radius, this implies that we want that:
\begin{align}
 & \left(\psi_{l}\left(S^{\mu}\right),S^{\mu}\frac{\partial}{\partial S^{\mu}}\psi_{l}\left(S^{\mu}\right)\right)\nonumber \\
 & =\left(a_{l}^{\mu},b_{l}^{\mu}\right)\left(\begin{array}{cc}
K_{l}^{\kappa}\left(S^{\mu}\right) & S^{\mu}\frac{\partial}{\partial S^{\mu}}K_{l}^{\kappa}\left(S^{\mu}\right)\\
J_{l}^{\kappa}\left(S^{\mu}\right) & S^{\mu}\frac{\partial}{\partial S^{\mu}}J_{l}^{\kappa}\left(S^{\mu}\right)
\end{array}\right)\label{eq:Continuity}
\end{align}
 Solving we get that :
\begin{equation}
\frac{b_{l}^{\mu}}{a_{l}^{\mu}}=-\frac{K_{l}^{\kappa}\left(S^{\mu}\right)}{J_{l}^{\kappa}\left(S^{\mu}\right)}\times\frac{\mathcal{D}_{K}\left(S^{\mu}\right)-\mathcal{D}_{\psi}\left(S^{\mu}\right)}{\mathcal{D}_{J}\left(S^{\mu}\right)-\mathcal{D}_{\psi}\left(S^{\mu}\right)}\label{eq:ratio-1}
\end{equation}
Where 
\begin{equation}
\mathcal{D}_{f}\left(S^{\mu}\right)=S^{\mu}\frac{f'\left(S^{\mu}\right)}{f\left(S^{\mu}\right)}\label{eq:Derivative-1}
\end{equation}
As such we obtain the KKR equation (see Eq. (\ref{eq:KKR_relation})): 
\begin{widetext}
\begin{equation}
\sum_{\mu lm}\left[B_{\nu\mu l'm';lm}\left(\mathbf{k}\right)+\frac{K_{l}^{\kappa}\left(S^{\mu}\right)}{J_{l}^{\kappa}\left(S^{\mu}\right)}\times\frac{\mathcal{D}_{K}\left(S^{\mu}\right)-\mathcal{D}_{\psi_{\mu}}\left(S^{\mu}\right)}{\mathcal{D}_{J}\left(S^{\mu}\right)-\mathcal{D}_{\psi_{\mu}}\left(S^{\mu}\right)}\delta_{\nu\mu;ll';mm'}\right]A_{lm}^{\mu}\left(\mathbf{k},E\right)=0\label{eq:KKR_general}
\end{equation}
\end{widetext}

\section{\protect\label{sec:Linearized-KKR}LKKR}

We now linearize KKR (LKKR). We write a new set of basis functions:
\begin{align}
 & \dot{a}_{l}^{\mu}\Phi_{lm}^{2\mu}\left(E,\mathbf{r},S^{\mu}\right)=Y_{lm}\left(\widehat{\mathbf{r-\mathbf{r}}_{\mu}}\right)\nonumber \\
 & \left\{ \begin{array}{cc}
\dot{\psi}_{l\mu}^{E}\left(\left|\mathbf{r}-\mathbf{r}_{\mu}\right|\right)-\dot{b}_{l}^{\mu}J_{l}^{\kappa}\left(\left|\mathbf{r}-\mathbf{r}_{\mu}\right|\right) & \left|\mathbf{r}-\mathbf{r}_{\mu}\right|\leq S^{\mu}\\
\dot{a}_{l}^{\mu}K_{l}^{\kappa}\left(\left|\mathbf{r}-\mathbf{r}_{\mu}\right|\right) & \left|\mathbf{r}-\mathbf{r}_{\mu}\right|>S^{\mu}
\end{array}\right.\label{eq:Energy_derivative_function}
\end{align}
For some linearization energy $E$. Furthermore we will assume that
\begin{align}
 & \Phi_{lm}^{1\mu}\left(E,\mathbf{r},S^{\mu}\right),\:\frac{\partial}{\partial\mathbf{r}}\Phi_{lm}^{1\mu}\left(E,\mathbf{r},S^{\mu}\right),\nonumber \\
 & \dot{\Phi}_{lm}^{2\mu}\left(E,\mathbf{r},S^{\mu}\right),\:\frac{\partial}{\partial\mathbf{r}}\dot{\Phi}_{lm}^{2\mu}\left(E,\mathbf{r},S^{\mu}\right)\label{eq:Continuous}
\end{align}
are continuous. We now look for wavefunctions of the form:
\begin{align}
\chi_{lm}^{\mu}\left(E,\mathbf{k},\mathbf{r}\right) & =\sum_{\mathbf{R}}\exp\left(i\mathbf{k}\cdot\mathbf{R}\right)\Phi_{lm}^{\mu}\left(E,\mathbf{r}-\mathbf{R},S^{\mu}\right)\nonumber \\
\dot{\chi}_{lm}^{\mu}\left(E,\mathbf{k},\mathbf{r}\right) & =\sum_{\mathbf{R}}\exp\left(i\mathbf{k}\cdot\mathbf{R}\right)\dot{\Phi}_{lm}^{\mu}\left(E,\mathbf{r}-\mathbf{R},S^{\mu}\right)\label{eq:Momentum_space}
\end{align}
 We now define the wavefunctions:
\begin{align}
\chi\left(E,\mathbf{k},\mathbf{r}\right) & =\sum_{\mu lm}A_{lm}^{\mu}\left(\mathbf{k},E\right)\chi_{lm}^{\mu}\left(E_{l}^{\mu},\mathbf{k},\mathbf{r}\right)\nonumber \\
 & +\sum_{\mu lm}B_{lm}^{\mu}\left(\mathbf{k},E\right)\dot{\chi}_{lm}^{\mu}\left(E_{l}^{\mu},\mathbf{k},\mathbf{r}\right)\label{eq:Linearization_wavefunction}
\end{align}
Where furthermore demand:
\begin{widetext}
\begin{align}
\sum_{\mu lm}\left[B_{\nu\mu l'm';lm}\left(\mathbf{k}\right)+\frac{K_{l}^{\kappa}\left(S^{\mu}\right)}{J_{l}^{\kappa}\left(S^{\mu}\right)}\times\frac{\mathcal{D}_{K}\left(S^{\mu}\right)-\mathcal{D}_{\psi_{\mu}}\left(S^{\mu}\right)}{\mathcal{D}_{J}\left(S^{\mu}\right)-\mathcal{D}_{\psi_{\mu}}\left(S^{\mu}\right)}\delta_{\nu\mu;ll';mm'}\right]A_{lm}^{\mu}\left(\mathbf{k}\right)\nonumber \\
+\sum_{\mu lm}\left[B_{\nu\mu l'm';lm}\left(\mathbf{k}\right)+\frac{K_{l}^{\kappa}\left(S^{\mu}\right)}{J_{l}^{\kappa}\left(S^{\mu}\right)}\times\frac{\mathcal{D}_{K}\left(S^{\mu}\right)-\mathcal{D}_{\dot{\psi}_{\mu}}\left(S^{\mu}\right)}{\mathcal{D}_{J}\left(S^{\mu}\right)-\mathcal{D}_{\dot{\psi}_{\mu}}\left(S^{\mu}\right)}\delta_{\nu\mu;ll';mm'}\right]B_{lm}^{\mu}\left(\mathbf{k}\right) & =0\label{eq:Linearized_KKR}
\end{align}
\end{widetext}

That is all the terms proportional to Bessel and Neumann Functions
vanish inside the spheres $\left|\mathbf{r}-\mathbf{r}_{\mu}\right|\leq S^{\mu}$
(see Eq. (\ref{eq:KKR_general}). Let us pick a basis of solutions
of Eq. (\ref{eq:Linearized_KKR})
\begin{equation}
\mathcal{A}_{1,\mu lm}^{n}\left(\mathbf{k}\right)=A_{lm}^{\mu}\left(\mathbf{k}\right),\:\mathcal{A}_{2,\mu lm}^{n}\left(\mathbf{k}\right)=B_{lm}^{\mu}\left(\mathbf{k}\right)\label{eq:Basis}
\end{equation}
 We now define the energy matrices to be: 
\begin{align}
\bar{H}_{\mathbf{k}}^{m,n} & =\sum_{\mu lm,\nu l'm'}\mathcal{A}_{i,\mu lm}^{m*}\left(\mathbf{k}\right)\left[-\bar{\Delta}^{\mathbf{k}}+\bar{V}_{KS}^{\mathbf{k}}\right]_{\mu lm,\nu l'm'}^{i,j}\mathcal{A}_{j,\nu l'm'}^{n}\left(\mathbf{k}\right)\nonumber \\
\bar{O}_{\mathbf{k}}^{m,n} & =\sum_{\mu lm,\nu l'm'}\mathcal{A}_{i,\mu lm}^{m*}\left(\mathbf{k}\right)\left[\bar{O}^{\mathbf{k}}\right]_{\mu lm,\nu l'm'}^{i,j}\mathcal{A}_{j,\nu l'm'}^{n}\left(\mathbf{k}\right)\label{eq:secular_matrices}
\end{align}
We now obtain the KS equation: 
\begin{equation}
\sum_{n}\bar{H}_{\mathbf{k}}^{m,n}V_{n}^{\mathfrak{A}}=\varepsilon^{\mathfrak{A}}\sum_{n}\bar{O}_{\mathbf{k}}^{m,n}V_{n}^{\mathfrak{A}}\label{eq:KS_equation}
\end{equation}

We have as such setup a LKKR Khon Sham like system for solving many
body DFT problems. Here $O$ is the overlap matrix.

\section{\protect\label{sec:LMTO}Equivalence of LKKR to LMTO}

\subsection{\protect\label{subsec:LMTO-review}LMTO review}

We note that the regular LMTO method is based on the following wavefunction
- which is a sum of three different wavefunctions given by:
\begin{widetext}
\begin{align}
 & \Theta_{lm}^{\mu}\left(E_{l}^{\mu},\mathbf{r}\right)=\Theta_{lm}^{1\mu}\left(E,\mathbf{r}\right)+\Theta_{lm}^{2\mu}\left(E,\mathbf{r}\right)+\Theta_{lm}^{3\mu}\left(E,\mathbf{r}\right)\nonumber \\
 & \Theta_{lm}^{1\mu}\left(E,\mathbf{r}\right)=\left\{ \begin{array}{cc}
Y_{lm}\left(\widehat{\mathbf{r-\mathbf{r}}_{\mu}}\right)\frac{K_{l}^{\kappa}\left(S^{\mu}\right)}{\Psi_{l}\left(\mathcal{D}_{K},S^{\mu}\right)}\Psi_{l}^{\mu}\left(\mathcal{D}_{K},\left|\mathbf{r}-\mathbf{r}_{\mu}\right|\right) & \left|\mathbf{r}-\mathbf{r}_{\mu}\right|\leq S^{\mu}\\
0 & otherwise
\end{array}\right.\nonumber \\
 & \Theta_{lm}^{2\mu}\left(E,\mathbf{r}\right)=\left\{ \begin{array}{cc}
\sum_{l'm'}Y_{l'm'}\left(\widehat{\mathbf{r-\mathbf{r}}_{\nu}}\right)\frac{J_{l'}^{\kappa}\left(S^{\nu}\right)}{\Psi_{l'}\left(\mathcal{D}_{J},S^{\nu}\right)}B_{\nu\mu l'm';lm}\left(\mathbf{k}\right)\Psi_{l'}^{\nu}\left(\mathcal{D}_{J},\left|\mathbf{r-\mathbf{r}}_{\nu}\right|\right) & \left|\mathbf{r}-\mathbf{r}_{\nu}\right|\leq S^{\nu}\\
0 & otherwise
\end{array}\right.\nonumber \\
 & \Theta_{lm}^{3\mu}\left(E,\mathbf{r}\right)=\left\{ \begin{array}{cc}
Y_{lm}\left(\widehat{\mathbf{r-\mathbf{r}}_{\mu}}\right)K_{l}^{\kappa}\left(\left|\mathbf{r}-\mathbf{r}_{\mu}\right|\right) & \left|\mathbf{r}-\mathbf{r}_{\nu}-\mathbf{R}\right|\geq S^{\nu},\forall\nu,\,\mathbf{R}\\
0 & otherwise
\end{array}\right.\label{eq:LMTO_basis}
\end{align}
\end{widetext}

Where 
\begin{align}
\Psi_{l}^{\mu}\left(\mathcal{D},\mathbf{r}\right) & =\psi_{l\mu}^{E}\left(r\right)+\omega_{l}\left(\mathcal{D}\right)\dot{\psi_{l\mu}^{E}}\nonumber \\
\omega_{l}\left(\mathcal{D}\right) & =-\frac{\psi_{l\mu}\left(S^{\mu}\right)}{\dot{\psi_{l\mu}}\left(S^{\mu}\right)}\cdot\frac{\mathcal{D-\mathcal{D}_{\psi}}}{\mathcal{D}-\mathcal{D}_{\dot{\psi}}}\label{eq:Definitions}
\end{align}
and 
\begin{equation}
\chi_{\mu,lm}^{MTO}\left(\mathbf{k}\right)=\sum_{\mathbf{R}}\exp\left(i\mathbf{k}\cdot\mathbf{R}\right)\Theta_{lm}^{\mu}\left(E_{l}^{\mu},\mathbf{r}-\mathbf{R},\mathbf{k}\right)\label{eq:MTO_wavefunction}
\end{equation}
is the LMTO basis set

\subsection{\protect\label{subsec:Equivalence-between-LKKR}Equivalence between
LKKR and LMTO calculation}

We wish to show that LKKR and LMTO are equivalent. To do se we now
need to check if:
\begin{align}
\chi_{\mu,lm}^{MTO}\left(\mathbf{k}\right) & \in Span\left\{ \chi_{l'm'}^{\mu}\left(E,\mathbf{r}\right),\dot{\chi}_{l'm'}^{\mu}\left(E,\mathbf{r}\right)\right\} \nonumber \\
\Rightarrow\chi_{\mu,lm}^{MTO}\left(\mathbf{k}\right) & =\sum_{\nu l'm'}A_{\nu l'm'}^{lm\Theta}\left(\mathbf{k}\right)\chi_{l'm'}^{\nu}\left(E,\mathbf{r}\right)+\nonumber \\
 & +\sum_{l'm'}B_{\nu l'm'}^{lm\Theta}\left(\mathbf{k}\right)\dot{\chi}_{l'm'}^{\nu}\left(E,\mathbf{r}\right)\label{eq:Span}
\end{align}
for some $A_{\nu l'm'}^{lm\Theta}\left(\mathbf{k}\right)$ and $B_{\nu l'm'}^{lm\Theta}\left(\mathbf{k}\right)$.
As such we would see that the LMTO method is spanned by the LKKR method,
however by counting the total number of basis set elements we would
then see the two are equivalent. However Eq. (\ref{eq:Span}) is a
direct check - see Appendix \ref{sec:Technical-Calculations}.

\section{\protect\label{sec:Linearized-KKR-1}HDKKR }

We now write the HDKKR wavefunctions. We write the following basis
set wavefunctions:
\begin{align}
 & \ddot{a}_{l}^{\mu}\Phi_{lm}^{3\mu}\left(E,\mathbf{r},S^{\mu}\right)=Y_{lm}\left(\widehat{\mathbf{r-\mathbf{r}}_{\mu}}\right)\times\nonumber \\
 & \times\left\{ \begin{array}{cc}
\ddot{\psi}_{l\mu}^{E}\left(\left|\mathbf{r}-\mathbf{r}_{\mu}\right|\right)-\ddot{b}_{l}^{\mu}J_{l}^{\kappa}\left(\left|\mathbf{r}-\mathbf{r}_{\mu}\right|\right) & \left|\mathbf{r}-\mathbf{r}_{\mu}\right|\leq S^{\mu}\\
\ddot{a}_{l}^{\mu}K_{l}^{\kappa}\left(\left|\mathbf{r}-\mathbf{r}_{\mu}\right|\right) & \left|\mathbf{r}-\mathbf{r}_{\mu}\right|>S^{\mu}
\end{array}\right.\label{eq:HDKKR_basis_III}
\end{align}
For some linearization energy $E$. and 
\begin{equation}
\Phi_{lm}^{i\mu}\left(E,\mathbf{r},S^{\mu}\right),\:\frac{\partial}{\partial\mathbf{r}}\Phi_{lm}^{i\mu}\left(E,\mathbf{r},S^{\mu}\right);\:i=1,2,3\label{eq:Continuous-1-1-1-1}
\end{equation}
Are continuous. We now define the wavefunctions:
\begin{equation}
\chi_{lm}^{i\mu}\left(E,\mathbf{k},\mathbf{r}\right)=\sum_{\mathbf{R}}\exp\left(i\mathbf{k}\cdot\mathbf{R}\right)\Phi_{lm}^{i\mu}\left(E_{l}^{\mu},\mathbf{r}-\mathbf{R},S^{\mu}\right)\label{eq:Basis-2-1}
\end{equation}
for $i=1,2,3$. We now look for wavefunctions of the form: 
\begin{equation}
\chi\left(E,\mathbf{k},\mathbf{r}\right)=\sum_{i\mu lm}A_{lm}^{i\mu}\left(\mathbf{k}\right)\chi_{lm}^{i\mu}\left(E_{l}^{\mu},\mathbf{k},\mathbf{r}\right)\label{eq:Linearization-1-1-1-1}
\end{equation}
Where furthermore demand:
\begin{widetext}
\begin{align}
\sum_{\mu lm}\left[B_{\nu\mu l'm';lm}\left(\mathbf{k}\right)+\frac{K_{l}^{\kappa}\left(S^{\mu}\right)}{J_{l}^{\kappa}\left(S^{\mu}\right)}\times\frac{\mathcal{D}_{K}\left(S^{\mu}\right)-\mathcal{D}_{\psi_{\mu}}\left(S^{\mu}\right)}{\mathcal{D}_{J}\left(S^{\mu}\right)-\mathcal{D}_{\psi_{\mu}}\left(S^{\mu}\right)}\delta_{\nu\mu;ll';mm'}\right]A_{lm}^{1\mu}\left(\mathbf{k}\right)\nonumber \\
+\sum_{\mu lm}\left[B_{\nu\mu l'm';lm}\left(\mathbf{k}\right)+\frac{K_{l}^{\kappa}\left(S^{\mu}\right)}{J_{l}^{\kappa}\left(S^{\mu}\right)}\times\frac{\mathcal{D}_{K}\left(S^{\mu}\right)-\mathcal{D}_{\dot{\psi}_{\mu}}\left(S^{\mu}\right)}{\mathcal{D}_{J}\left(S^{\mu}\right)-\mathcal{D}_{\dot{\psi}_{\mu}}\left(S^{\mu}\right)}\delta_{\nu\mu;ll';mm'}\right]A_{lm}^{2\mu}\left(\mathbf{k}\right)\nonumber \\
+\sum_{\mu lm}\left[B_{\nu\mu l'm';lm}\left(\mathbf{k}\right)+\frac{K_{l}^{\kappa}\left(S^{\mu}\right)}{J_{l}^{\kappa}\left(S^{\mu}\right)}\times\frac{\mathcal{D}_{K}\left(S^{\mu}\right)-\mathcal{D}_{\ddot{\psi}_{\mu}}\left(S^{\mu}\right)}{\mathcal{D}_{J}\left(S^{\mu}\right)-\mathcal{D}_{\ddot{\psi}_{\mu}}\left(S^{\mu}\right)}\delta_{\nu\mu;ll';mm'}\right]A_{lm}^{3\mu}\left(\mathbf{k}\right) & =0\label{eq:HDKKR_main_equation}
\end{align}
\end{widetext}

So that Bessel like terms vanish inside the MT spheres (see Eqs. (\ref{eq:Linearized_KKR})
and (\ref{eq:KKR_general})). Let us pick a basis of solutions of
Eq. (\ref{eq:HDKKR_main_equation}):
\begin{equation}
\mathcal{A}_{i,\mu lm}^{n}\left(\mathbf{k}\right)=A_{lm}^{i\mu}\left(\mathbf{k}\right)\label{eq:Basis-1}
\end{equation}
 We now use Eqs. (\ref{eq:secular_matrices}) and (\ref{eq:KS_equation})
to complete calculations.

\section{\protect\label{sec:HDMTO}HDMTO}

The HDMTO basis set is based on the same wavefunctions as in Eq. (\ref{eq:LMTO_basis})
where however:

\begin{equation}
\Psi_{l}^{\mu}\left(\mathcal{D},\mathbf{r}\right)=\psi_{l\mu}^{E}\left(r\right)+\dot{\omega}_{l}^{\mu}\left(\mathcal{D},\mathcal{D}^{2}\right)\dot{\psi_{l\mu}^{E}}+\ddot{\omega}_{l}^{\mu}\left(\mathcal{D},\mathcal{D}^{2}\right)\ddot{\psi}_{l\mu}^{E}\label{eq:Radial_wavefunction-1}
\end{equation}
Where 
\begin{equation}
\mathcal{D}_{f}\left(S\right)=S^{2}\frac{f"\left(S\right)}{f\left(S\right)}\label{eq:Derivative-1-1}
\end{equation}
We want that: 
\begin{align}
 & \frac{K_{l}^{\kappa}\left(S^{\mu}\right)}{\Psi_{l}\left(\mathcal{D}_{K},S^{\mu}\right)}\left(\begin{array}{c}
\Psi_{l}\left(S^{\mu}\right)\\
S^{\mu}\frac{\partial}{\partial S^{\mu}}\Psi_{l}\left(S^{\mu}\right)\\
\left(S^{\mu}\right)^{2}\frac{\partial^{2}}{\partial\left(S^{\mu}\right)^{2}}\Psi_{l}\left(S^{\mu}\right)
\end{array}\right)\nonumber \\
 & =\left(\begin{array}{c}
K_{l}^{\kappa}\left(S^{\mu}\right)\\
S^{\mu}\frac{\partial}{\partial S^{\mu}}K_{l}^{\kappa}\left(S^{\mu}\right)\\
\left(S^{\mu}\right)^{2}\frac{\partial^{2}}{\partial\left(S^{\mu}\right)^{2}}K_{l}^{\kappa}\left(S^{\mu}\right)
\end{array}\right)\label{eq:Continuity-1}
\end{align}
so that the wavefunction along with its first and second derivatives
are continuous. The derivation is identical to the one in Appendix
\ref{sec:Technical-Calculations}. Now we write:
\begin{equation}
\chi_{\mu,lm}^{MTO}\left(\mathbf{k}\right)=\sum_{\mathbf{R}}\exp\left(i\mathbf{k}\cdot\mathbf{R}\right)\Theta_{lm}^{\mu}\left(E,\mathbf{r}-\mathbf{R},\mathbf{k}\right)\label{eq:MTO_wavefunction-1}
\end{equation}
And study this basis set. 

\subsection{\protect\label{subsec:Equivalence-of-HDMTO}Differences and similarities
between HDMTO and HDKKR}

We note that HDKKR basis set includes HDMTO as HDKKR only demands
continuity of the wavefunction and its derivative while HDMTO also
demands continuity of the second derivative. Furthermore because of
the similarities of the form of the HDMTO and LMTO basis wavefunctions
the HDMTO basis set eliminates all Bessel like wavefunction components
inside all MT spheres so HDMTO basis wavefunctions are a type of HDKKR
basis wavefunctions. As such HDKKR has greater variational power then
HDMTO. We note that this observation is true, despite the fact that
the exact solution of the KS Hamiltonian have all derivatives continuous,
as wavefunctions with discontinuous second derivatives may help approximate
ones with continuous second derivatives better. As the KS Equations
are variational HDKKR gives a better estimate of the ground state
energy then HDMTO. However the HDMTO method is much more efficient
as the basis is smaller and it does not require solving Eq. (\ref{eq:HDKKR_main_equation})
on computer as the basis set automatically satisfies it as there are
no Bessel like functions inside any MT spheres so in many case HDMTO
is preferable to HDKKR.

\section{\protect\label{sec:Conclusion}Conclusion}

In this work we have studied a linearized version of KKR (LKKR) and
shown it to be exactly equivalent to LMTO. We have extended these
ideas to higher derivative HDKKR and HDMTO basis sets. These two basis
sets are not equivalent. The HDKKR basis set has greater variational
power then HDMTO basis set as it includes it as a subset. However
HDMTO is much easier to work with, requires significantly less computational
power and no auxiliary equations to solve. As such in many cases,
since HDMTO is easier to implement, it is likely the preferred basis
set method. In the future it would be of interest to set up practical
HDMTO calculations for real solid state crystal systems.

\appendix

\section{\protect\label{sec:Technical-Calculations}Technical Calculations}

As such for the interstitial region to match between LKKR and LMTO
we must have that: 
\begin{equation}
A_{\nu l'm'}^{lm\Theta}\left(\mathbf{k}\right)+B_{\nu l'm'}^{lm\Theta}\left(\mathbf{k}\right)=\delta_{\mu\nu}\delta_{l,l';m,m'}\label{eq:Delta_function}
\end{equation}
Now we match the MT region. From which we read of: 
\begin{widetext}
\begin{align}
A_{\nu l'm'}^{lm\mu\Theta}\left(\mathbf{k}\right) & =\delta_{\mu,\nu;l,l';m,m'}\frac{K_{l}^{\kappa}\left(S^{\mu}\right)}{\Psi_{l}\left(\mathcal{D}_{K},S^{\mu}\right)}a_{l}^{\mu}+\frac{J_{l'}^{\kappa}\left(S^{\mu}\right)}{\Psi_{l'}\left(\mathcal{D}_{J},S^{\mu}\right)}B_{\nu\mu l'm';lm}\left(\mathbf{k}\right)a_{l'}^{\nu}\nonumber \\
B_{\nu l'm'}^{lm\mu\Theta}\left(\mathbf{k}\right) & =\delta_{\mu,\nu;l,l';m,m'}\frac{K_{l}^{\kappa}\left(S^{\mu}\right)}{\Psi_{l}\left(\mathcal{D}_{K},S^{\mu}\right)}\omega_{l}\left(\mathcal{D}_{K}\right)\dot{a}_{l}^{\mu}+\frac{J_{l'}^{\kappa}\left(S^{\mu}\right)}{\Psi_{l'}\left(\mathcal{D}_{J},S^{\mu}\right)}B_{\nu\mu l'm';lm}\left(\mathbf{k}\right)\omega_{l'}\left(\mathcal{D}_{J}\right)\dot{a}_{l'}^{\nu}\label{eq:Solution-1}
\end{align}
So we want: 
\[
\delta_{\mu,\nu;l,l';m,m'}\frac{K_{l}^{\kappa}\left(S^{\mu}\right)}{\Psi_{l}\left(\mathcal{D}_{K},S^{\mu}\right)}\left[a_{l}^{\mu}+\omega_{l}\left(\mathcal{D}_{K}\right)\dot{a}_{l}^{\mu}\right]+B_{\nu\mu l'm';lm}\left(\mathbf{k}\right)\frac{J_{l'}^{\kappa}\left(S^{\mu}\right)}{\Psi_{l'}\left(\mathcal{D}_{J},S^{\mu}\right)}\left[a_{l'}^{\nu}+\omega_{l'}\left(\mathcal{D}_{J}\right)\dot{a}_{l'}^{\nu}\right]=\delta_{\mu\nu}\delta_{l,l';m,m'}
\]
Now we have that:

\begin{align}
 & a_{l}^{\mu}+\omega_{l}\left(\mathcal{D}_{J}\right)\dot{a}_{l}^{\mu}\nonumber \\
 & =a_{l}^{\mu}-\frac{\psi_{l\mu}\left(S^{\mu}\right)}{\dot{\psi_{l\mu}}\left(S^{\mu}\right)}\cdot\frac{\mathcal{D}_{J}\mathcal{-\mathcal{D}_{\psi}}}{\mathcal{D}_{J}-\mathcal{D}_{\dot{\psi}}}\dot{a}_{l}^{\mu}\nonumber \\
 & =S^{\mu}\psi_{l\mu}\left(S^{\mu}\right)J_{l}^{\kappa}\left(S^{\mu}\right)\left(\mathcal{D}_{J}-\mathcal{D}_{\psi}\right)-\frac{\psi_{l\mu}\left(S^{\mu}\right)}{\dot{\psi_{l\mu}}\left(S^{\mu}\right)}\cdot\frac{\mathcal{D}_{J}\mathcal{-\mathcal{D}_{\psi}}}{\mathcal{D}_{J}-\mathcal{D}_{\dot{\psi}}}S^{\mu}\dot{\psi}_{l\mu}\left(S^{\mu}\right)J_{l}^{\kappa}\left(S^{\mu}\right)\left(\mathcal{D}_{J}-\mathcal{D}_{\dot{\psi}}\right)=0\label{eq:Check}
\end{align}
As such Eq. (\ref{eq:Delta_function}) is satisfied. Next we wish
to show that the continuity equation is satisfied for LMTO basis wavefunctions.
Indeed: 
\begin{align}
 & \frac{K_{l}^{\kappa}\left(S^{\mu}\right)}{\Psi_{l}\left(\mathcal{D}_{K},S^{\mu}\right)}\left[a_{l}^{\mu}+\omega_{l}\left(\mathcal{D}_{K}\right)\dot{a}_{l}^{\mu}\right]\nonumber \\
 & =\frac{K_{l}^{\kappa}\left(S^{\mu}\right)}{\Psi_{l}\left(\mathcal{D}_{K},S^{\mu}\right)}\left[a_{l}^{\mu}+\frac{\psi_{l\mu}\left(S^{\mu}\right)}{\dot{\psi_{l\mu}}\left(S^{\mu}\right)}\cdot\frac{\mathcal{D}_{K}\mathcal{-\mathcal{D}_{\psi}}}{\mathcal{D}_{K}-\mathcal{D}_{\dot{\psi}}}\dot{a}_{l}^{\mu}\right]\nonumber \\
 & =\frac{K_{l}^{\kappa}\left(S^{\mu}\right)}{\psi_{l\mu}\left(S^{\mu}\right)\left[1-\frac{\mathcal{D}_{K}-\mathcal{D}_{\psi}}{\mathcal{D}_{K}-\mathcal{D}_{\dot{\psi}}}\right]}\left[S^{\mu}\psi_{l\mu}\left(S^{\mu}\right)J_{l}^{\kappa}\left(S^{\mu}\right)\left(\mathcal{D}_{J}-\mathcal{D}_{\psi}\right)+\frac{\psi_{l\mu}\left(S^{\mu}\right)}{\dot{\psi_{l\mu}}\left(S^{\mu}\right)}\cdot\frac{\mathcal{D}_{K}\mathcal{-\mathcal{D}_{\psi}}}{\mathcal{D}_{K}-\mathcal{D}_{\dot{\psi}}}\left[S^{\mu}\dot{\psi}_{l}\left(S^{\mu}\right)J_{l}^{\kappa}\left(S^{\mu}\right)\left(\mathcal{D}_{J}-\mathcal{D}_{\cdot\psi}\right)\right]\right]\nonumber \\
 & =\frac{S^{\mu}K_{l}^{\kappa}\left(S^{\mu}\right)J_{l}^{\kappa}\left(S^{\mu}\right)}{\mathcal{D}_{\psi}-\mathcal{D}_{\dot{\psi}}}\left[\left[\mathcal{D}_{K}-\mathcal{D}_{\dot{\psi}}\right]\left(\mathcal{D}_{J}-\mathcal{D}_{\psi}\right)-\left[\mathcal{D}_{K}\mathcal{-\mathcal{D}_{\psi}}\right]\left(\mathcal{D}_{J}-\mathcal{D}_{\dot{\psi}}\right)\right]\nonumber \\
 & =S^{\mu}K_{l}^{\kappa}\left(S^{\mu}\right)J_{l}^{\kappa}\left(S^{\mu}\right)\left[\mathcal{D}_{K}-\mathcal{D}_{J}\right]=1\label{eq:One}
\end{align}
As such Eq. (\ref{eq:Linearized_KKR}) is verified for the LMTO basis
set as it is a continuous continuously differentiable wavefunction
in the span of the KKR basis set with no components of the form of
Bessel functions inside the MT spheres.
\end{widetext}

\end{document}